\documentstyle[aps,twocolumn,10pt]{revtex}
\newlength{\defaultparindent}
\newenvironment{Default Paragraph Font}{}{}

\begin{document}
\title{Optical properties of tungsten thin films perforated with a bidimensional
array of subwavelength holes}
\author{Micha\"{e}l SARRAZIN, Jean-Pol VIGNERON}
\address{Laboratoire de Physique du Solide\\
Facult\'{e}s Universitaires Notre-Dame de la Paix\\
Rue de Bruxelles 61, B-5000 Namur, Belgium}
\maketitle

\begin{abstract}
We present a theorical investigation of the optical transmission of a
dielectric grating carved in a tungsten layer. For appropriate wavelengths
tungsten shows indeed a dielectric behaviour. Our numerical simulations
leads to theoretical results similar to those found with metallic systems
studied in earlier works. The interpretation of our results rests on the
idea that the transmission is correlated with the resonant response of
eigenmodes coupled to evanescent diffraction orders.
\end{abstract}

For a few years, properties and technological applications of one- or
two-dimensionnal metallic gratings have received a growing interest. In
1998, Ebbesen {\it et} {\it al} [1] reported on optical transmission
experiments performed on periodic arrays of subwavelength cylindrical holes
drilled in a thin metallic layer deposited on glass. These experiments
renewed the motivation for investigating metallic gratings. Two attractive
characteristics of their results are often cited : the transmission, which
is a lot higher than the addition of individual holes contributions, and the
peculiar wavelength dependance of the transmission. Further work [1-7] has
suggested that these features arise from the presence of the metallic layer,
and call for the presence of surface plasmons in order to explain these
transmission characteristics. In particular, they have identified the convex
high transmission regions, i.e., the regions between the minima, as regions
dominated by the plasmon response. However, many questions remain and need
to be answered in order to clarify the mechanisms involved in these
experiments.

In a recent paper [8], extensive simulations have been performed in order to
understand the optical properties of a chromium layer similar to those
developed in experiments. Recalling of Wood's anomalies [9], it is shown
that the transmission and reflection are better described as Fano's profiles
correlated with resonant response of the eigenmodes coupled to
nonhomogeneous diffraction orders. Indeed, as explained by V.U. Fano [10],
A. Hessel and A.A. Oliner [11] for one dimensional gratings, Wood's
anomalies are related to eigenmodes grating excitation. To be accurate, they
have shown that Wood's anomalies [11] may arise in two ways. The first case
occurs at Rayleigh's wavelengths, when a diffracted order becomes grazing to
the grating plane [12]. The diffracted beam intensity then increases just
before the diffracted order vanishes. The other case is related to a
resonance effect. Such resonances come from coupling between the incident
light and the eigenmodes of the grating. Both types of anomalies may occur
separately and independently, or may almost coincide.

In our previous paper [8], as we used metal in our device, it seemed natural
to assume that these resonances are surface plasmons. Nevertheless, it is
important to note that our analysis did not make any hypothesis on the
origin of the eigenmodes. This implies that it could be possible to obtain
transmission curves similar to those found for metals, by substituting
guided modes or other types of polaritons to the surface plasmons. This
would allow, for instance, to substitute dielectric guided modes to metallic
excitations. This is the purpose of the present paper. We could also use
films made of ionic crystal in the far infrared, and deal with
phonon-polaritons (in progress).

In this paper we perform simulations of a device which consists of arrays of
subwavelength cylindrical holes in a tungsten layer deposited on glass
substrate (figure 1). Indeed, tungsten becomes dielectric on a restricted
domain of wavelength in the range $240-920$ $nm,$ i.e. the real part of its
permittivity becomes positive. The permittivity values are taken from
experiments [13]. So, whereas plasmons cannot exist, we show that the
transmission pattern is similar to that obtained in the case of a metallic
film. However, experiment has also shown that a germanium film is not able
to give rise to an Ebbesen effect. We then also need to explain here why in
the case of germanium nothing interesting can be observed.

Our simulations rest on a coupled modes method (which takes into account the
periodicity of the device permittivity) associated with the use of the
scattering matrix formalism [8,14]. In this way, we calculate the amplitudes
of the reflected and transmitted field, for each diffracted order (which
correspond to a vector $\overrightarrow{g}$ of the reciprocal lattice)
according to their polarization ($s$ or $p$). In the following, for a square
grating of parameter $a$, note that, $\overrightarrow{g}=\frac{2\pi }a(i,$ $%
j)$, so that the pair of integers $(i,$ $j)$ denotes the corresponding
vector of the reciprocal lattice (i.e. diffraction order). In addition, we
recall that Rayleigh's wavelength are defined as 
\begin{equation}
\lambda _R^{u,i,j}=a\sqrt{\varepsilon _u}(i^2+j^2)^{-\frac 12}
\end{equation}
where $\varepsilon _u$\ represents either the permittivity of the vacuum ($%
\varepsilon _v$), or of the dielectric substrate ($\varepsilon _d$). We
calculate the zero transmission order, referring to the experimental
measurements performed in the metallic case. We can also estimate the
partial density of states (PDOS) for some positions in the first Brillouin's
zone.

The calculated transmission of the incident wave is shown against wavelength
in Fig.\ 2a for the zero diffraction order, and for incident ligth normal to
the surface of a W film on glass. The holes diameter is set to $d=320$ $nm$
and the thickness of the film is $h=100$ $nm$. The transmission is given for
square gratings of parameter $a=550$ $nm$, $500$ $nm$ and $450$ $nm$,
respectively. In Fig.\ 2a, it is shown that the transmission increases with
the wavelength, and that it is characterized by minima marked 1 and 2 on the
figure, which are located closer after Rayleigh's wavelength $\lambda
_R^{d,0,\pm 1}$ and $\lambda _R^{v,0,\pm 1}$, respectively. The locations of
these minima are shifted toward larger wavelengths when the grating size
increases. Our investigations show that these minima disappear when
considering a system without hole. Note that these results are qualitatively
similar to those from the experimental data of Ebbesen {\it et} {\it al} in
the case of metallic layers [1,2].

Fig 2b shows the calculated zero-order transmission as a function of the
incident wavelength, for gratings with differents holes diameters. As in the
case of metallic gratings [1,2,6,7], it is clear that transmission curves
does not depend of the hole diameter except for the amplitude of the
spectral features.

In order to underline the correlation between the transmission amplitude and
the holes diameter we also give in Fig. 2c the transmission amplitude for
many hole diameter values (dots in Fig. 2c). The transmission values are
those which correspond to the maximum located at $576.7$ $nm$ for a square
grating of parameter $a=500$ $nm$, with the chosen thickness of the film ($%
h=100$ $nm$). We show that the transmission amplitude increases
exponentially as a function of the hole diameter (solid line in Fig. 2c).
This result suggests that there is no cavity mode involved in those
phenomena.

In Fig 2d we give the calculated transmission as a function of the
wavelength of the incident wave for the zeroth diffraction order, for
different layer thicknesses. As in the case of metallic gratings [1,2,6,7],
we show that the behaviour of the transmission curves does not depend of the
thickness except for the overall amplitude.

The results we shown in Fig 2 present behaviours very similar to those
obtained with metals cases [1,2], even though we use tungsten in its
dielectric domain. Let us discuss this point. As explained in [8], the
eigenmodes of the grating play a crucial role in those experiments. Let us
first recall that, reflected and transmitted amplitudes are linked to the
incident field through the $S$ scattering\ matrix.\ Let us define $F_{scat}$
as the scattered field, and $F_{in}$ as the incident field. Then, $F_{scat}$%
\ is related to $F_{in}$ {\it via} the scattering matrix, in such a way that 
\begin{equation}
S(\lambda )F_{in}(\lambda )=F_{scat}(\lambda )
\end{equation}
In this way, the eingenmodes of the structure are solution of eq.2 which
exist in absence of source, i.e. when 
\begin{equation}
S^{-1}(\lambda _\eta )F_{scat}(\lambda _\eta )=0
\end{equation}
This homogeneous problem is well know in the theory of gratings [8,11].
Complex wavelengths $\lambda _\eta =\lambda _\eta ^R+i\lambda _\eta ^I$, for
which eq.3 has non-trivial solutions require that 
\begin{equation}
\det (S^{-1}(\lambda _\eta ))=0
\end{equation}
so that they coincide with the poles of the complex function $\det
(S(\lambda ))$.

If we extract the singular part of $S$ corresponding to the eigenmodes of
the structure, we can write $S$ as [8,11] 
\begin{equation}
S(\lambda )=\sum_\eta \frac{R_\eta }{\lambda -\lambda _\eta }+S_h(\lambda )
\end{equation}
i.e. under the form of a generalized Laurent series, where $R_\eta $ is the
residue associated which the pole $\lambda _\eta $ and $S_h(\lambda )$ is
the holomorphic part of $S$ which corresponds to physically non-resonant
process.

Thus, assuming that $f(\lambda )$ is the $m^{th}$ component of $%
F_{scatt}(\lambda ),$ we have, for the expression of $f(\lambda )$ in the
neighboorhood of one pole $\lambda _\eta $ [8,10,11] 
\begin{equation}
f(\lambda )=\frac{r_\eta }{\lambda -\lambda _\eta }+s(\lambda )
\end{equation}
where $r_\eta =$ $\left[ R_\eta F_{in}\right] _m$ and $s(\lambda )=\left[
S_h(\lambda )F_{in}\right] _m$.

We consider the case where non-resonant processes cannot be neglected, and
assume that $s(\lambda )\sim s_0$ is a constant value. Thus, it is easy to
show that eq.6 can be written as [8,11] 
\begin{equation}
\left| f(\lambda )\right| ^2=\frac{\left( \lambda -\lambda _z^R\right)
^2+\lambda _z^{I\ \ 2}}{\left( \lambda -\lambda _\eta ^R\right) ^2+\lambda
_\eta ^{I\ \ 2}}\left| s_0\right| ^2
\end{equation}
with 
\begin{equation}
\lambda _z^R=\lambda _\eta ^R-\nu ^R\text{ \ \ \ and \ \ }\lambda
_z^I=\lambda _\eta ^I-\nu ^I
\end{equation}
where 
\begin{equation}
\nu =\frac{r_\eta }{s_0}
\end{equation}
Coefficient $\nu $ measures the significance of the resonant effect,
compared with non-resonant contributions. $\lambda _z=\lambda _z^R+i\lambda
_z^I${\it \ }denotes the zero of eq.6 and 7. This last expression take into
account the interferences between resonant and non-resonant processes, which
lead to asymmetric transmission profiles. Eq.7 leads to a typically resonant
process (described by a lorentzian curve) or to a typical asymmetric
behavior where a minimum is following a maximum, and {\it vice versa}
assuming the values of $\nu $. We note that this properties, which results
from the interference between resonant and non-resonant processes, are
similar to those described by A.A. Hessel, A. Oliner [11] and V.U. Fano
[10]. The profiles like those describe on figure 3a are often reffered to as
''Fano's profiles''.

It is now necessary to find the structure eigenmodes. To this purpose, we
show in Fig. 3b the partial density of states, i.e. the density of states in
the center $\Gamma $ of the first Brillouin's zone. Indeed, when the
incident light is normal to the surface, our assumption is that the DOS at $%
\Gamma $ is the dominant contribution to those phenomena.

The solid line shows the partial DOS which is calculated in the case of a
grating with $d=320$ $nm$, $h=100$ $nm$ and $a=500$ $nm$. The dashed line
shows the partial DOS which is calculated for a similar system in absence of
holes. The overall pattern of the partial DOS of the grating is related to
the pattern of the partial DOS of the planar tungsten layer. It appears some
sharp localized peaks related to the eigenmodes of the tungsten grating.
Such eigenmodes can only be associated with guided modes of the tungsten
grating assuming its dielectric properties. We note the Rayleigh's anomalies
(circles around sharp minima of DOS) which appear just on the side of the
eigenmodes at shorter wavelength. We also show the corresponding
transmission for sake of comparison. One clearly notices the absence of
coincidence between the position of the eigenmodes and the transmission
maxima. Nevertheless the eigenmodes are behind of the typical profiles of
the transmission curves [8]. In fact, one can interpret the features
lineshapes as arising from resonant Wood's anomalies, similar to those
studied by V. U. Fano [10] and by A. Hessel and A.A. Oliner [11].

Let us consider an incident propagative wave, which diffracts and generates
a evanescent diffraction order. Such an order is coupled with a guided
eigenmode which is characterized by a complex wavelength $\lambda _\eta $.
It becomes possible to excite this eigenmode which lead to a feedback
reaction on the evanescent order. For this reason, A. Hessel and A.A. Oliner
called such order a ''resonant order''. This process is related to the
resonant term.

The evanescent resonant order can diffract due to the layer corrugation and
generates a contribution to the propagative zero diffraction order. Thus,
one can ideally expect to observe a resonant (lorentzian) profile for the
zero diffraction order. Nevertheless, it is necessary to take into account a
nonresonant diffraction process related to the holomorphic term of the
scattering matrix. So, the incident wave, generates a propagative zero
order. Then, one takes into account the interference of two rates, i.e.
resonant and non resonant contribution to order zero. It appears profiles
which are typically the Fano's profiles which correspond to resonant process
where one takes into account nonresonant effects. One note that a maximum in
transmission does not necessary correspond to the maximum of resonance of a
diffraction order. So, the Fano's profiles behaviour of the transmission,
result from the superimposing of resonant and nonresonant contributions to
the zero diffraction order.

If one refers to the present situation the resonance is close to Rayleigh's
wavelength. Consequently, asymmetric Fano's profiles are shifted toward
Rayleigh's wavelength in the same way. Then the transmission minima, which
in fact correspond to minima of Fano's profiles, disappears when crossing a
Rayleigh's wavelength. Maxima of the transmission, which is shifted towards
larger wavelengths as shown in [8], correspond to the maxima of Fano's
profiles.

Then, the maxima do not correspond explicitly to maxima of resonances but
rather to the maxima of Fano's profiles originating from the excitation of
guided-modes and not surface plasmons as in metallic cases [8].
Nevertheless, it is necessary to emphasize why in the case of germanium,
which is a dielectric, nothing similar has been observed. In fact, in the
wavelength domain of our simulations the imaginary part of the tungsten
permittivity is large, whereas in the case of germanium, in the domain
wavelength studied by Ebbesen {\it et al}, the imaginary part of the
permittivity is weak enough to consider germanium as transparent. In such
situation, the contribution of the holomorphic part of the scattering matrix
is much larger than the resonant part so that the latter can be neglected.
Thus one completely loses the benefit of the resonant effects which plays a
fundamental role in the transmission lineshapes. Morevover, in the case of
tungsten, the opacity is such that there is not enough direct transmitted
light to mask the effects of resonant processes.

It would be desirable to obtain more experimental data about tungsten
gratings, as a way to confirm our theorical findings. The ability to use
dielectric layers would open a potential of Ebbesen's devices. One notes
that the role of guided resonances and Fano's profiles in photonic crystal
slabs have been also underlined by S. Fan and J. D. Joannopoulos in a recent
paper [15].

\section*{Acknowledgments}

The authors thank J.-M. Vigoureux and D. Van Labeke for their invaluable
advice and useful discussions.

We acknowledge the use of Namur Scientific Computing Facility (Namur-SCF), a
common project between the FNRS, IBM Belgium, and the Facult\'{e}s
Universitaires Notre-Dame de la Paix (FUNDP).

This work was carried out with support from EU5 Centre of Excellence
ICAI-CT-2000-70029 and from the Inter-University Attraction Pole (IUAP P5/1)
on ''Quantum-size effects in nanostructured materials'' of the Belgian
Office for Scientific, Technical, and Cultural Affairs.

\section*{References}

[1] T.W. Ebbesen, H.J. Lezec, H.F. Ghaemi, T. Thio, P.A. Wolff, Nature
(London) 391, 667 (1998)

[2] H.F. Ghaemi, T. Thio, D.E. Grupp, T.W. Ebbesen, H.J. Lezec, Phys. Rev.
B, 58, 6779 (1998)

[3] U. Schr\"{o}ter, D. Heitmann, Phys. Rev. B, 58, 15419 (1998)

[4] J.A. Porto, F.J. Garcia-Vidal, J.B. Pendry, Phys. Rev. Lett., 83, 2845
(1999)

[5] D.E. Grupp, H.J. Lezec, T.W. Ebbesen, K.M. Pellerin, T. Thio, Appl.
Phys. Lett. 77 (11) 1569 (2000)

[6] T. Thio, H.J. Lezec, T.W. Ebbesen, Physica B 279, 90 (2000)

[7] A. Krishnan, T. Thio, T. J. Kim, H. J. Lezec, T. W. Ebbesen, P.A. Wolff,
J.\ Pendry, L. Martin-Moreno, F. J. Garcia-Vidal, Opt. Commun., 200, 1 (2001)

[8] M. Sarrazin, J.-P. Vigneron, J.-M. Vigoureux, Phys. Rev. B, 67, 085415
(2003)

[9] R.W. Wood, Phys. Rev. 48, 928 (1935)

[10] V.U. Fano, Ann. Phys. 32, 393 (1938)

[11] A. Hessel, A. A. Oliner, Appl. Opt. 4 (10) 1275 (1965)

[12] Lord Rayleigh, Proc. Roy. Soc. (London) A79, 399 (1907)

[13] D.W. Lynch, W.R. Hunter, in Handbook of Optical Constants of Solids II,
edited by E.D. Palik (Academic Press, New York, 1991)

[14] J.P. Vigneron, F. Forati, D. Andr\'{e}, A. Castiaux, I. Derycke, A.
Dereux, Ultramicroscopy, 61, 21 (1995)

[15] S. Fan, J. D. Joannopoulos, Phys. Rev. B, 65, 235112 (2002)

\section*{Captions}

FIG. 1. (a) Diagrammatic view of the system unders study. Transmission is
calculated for the zeroth order and at normal incidence as in experiments.
(b) Real and imaginary part of tungsten permittivity.

FIG. 2. Percentage transmission of the incident wave on the surface against
wavelength, for the zeroth diffraction order, for differents parameters of
the square grating (a), for differents holes diameters (b) and for different
thickness (d). Part (c) represent the transmission against hole diameter for
a given wavelength.

FIG. 3. (a) Representation of typical Fano's profiles. (b) PDOS\ against
wavelength compare with transmission.

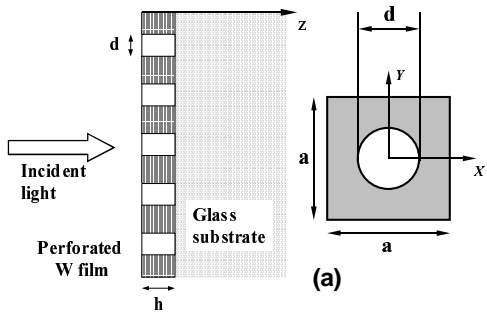
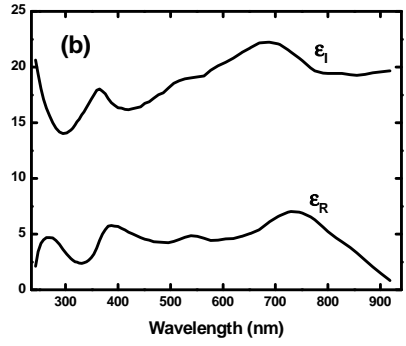
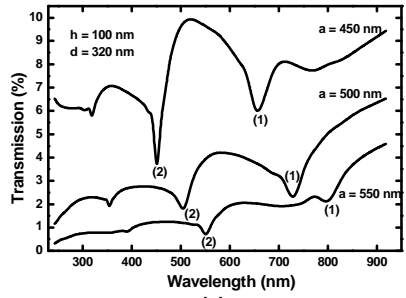
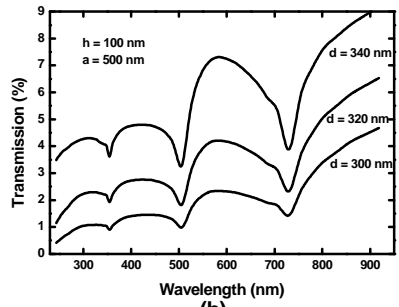
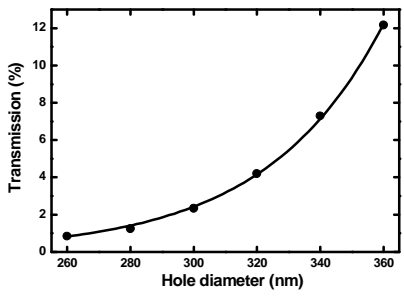
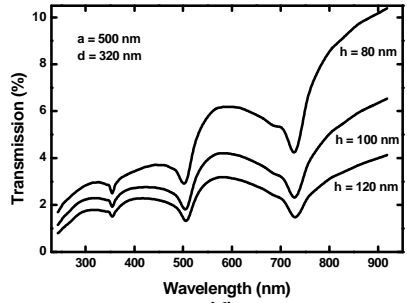
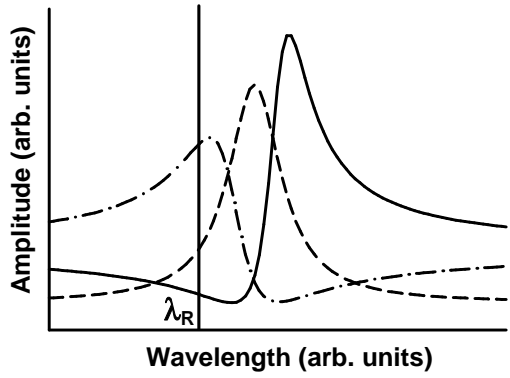
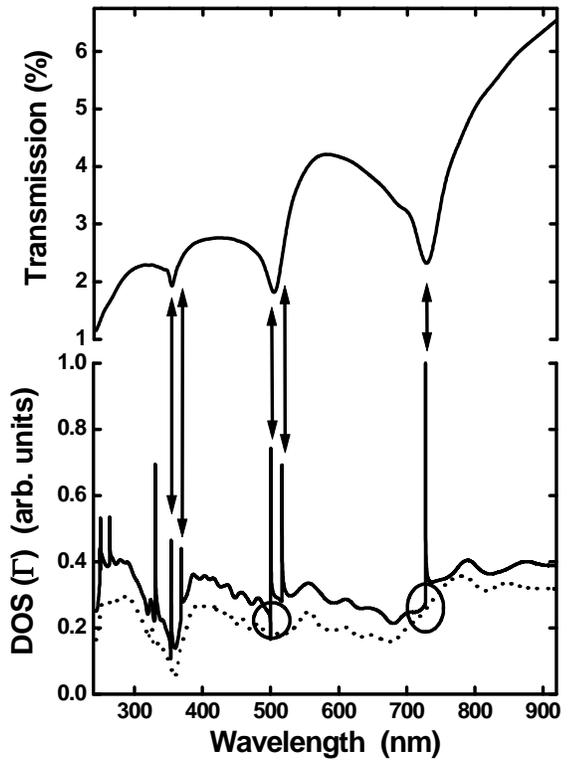
\end{document}